\documentclass{ws-procs9x6}
\def\Journal#1#2#3#4{{#4}, {#1}, {#2}, #3} 
\newcommand{\etal}{et alii}
\usepackage{verbatim}
\begin{document}

\title{Light Nuclei and Isotope Abundances in Cosmic Rays. \\Results from AMS-01}
\author{N. Tomassetti$^{*}$, for the AMS-01 Collaboration}
\address{Perugia University and INFN Perugia,\\
  Perugia, I-06122, Italy\\
$^{*}$E-mail: nicola.tomassetti@pg.infn.it\\
www.pg.infn.it}

\begin{abstract} 
 Observations of the chemical and isotopic composition of light cosmic-ray nuclei can be used to constrain the propagation models. Nearly 200,000 light nuclei ($Z>2$) have been observed by AMS-01 during the 10-day flight STS-91 in June 1998. Using these data, we have measured Li, Be, B and C in the kinetic energy range 0.35 \-- 45 GeV/nucleon. In this proceeding, our charge and isotopic composition results are presented and discussed.
\end{abstract}

\bodymatter
\section{Introduction} 
Cosmic rays detected on Earth with kinetic energies from 100 MeV to 100 GeV per nucleon
are believed to be produced by particle acceleration mechanisms occurring
in galactic sites such as supernova remnants. 
Most of our knowledge on the insterstellar propagation of the galactic cosmic rays
comes from the study of secondary species, i.e. spallation products of C-N-O and Fe nuclei
that are almost absent in the cosmic ray (CR) sources.

The relation between secondary CRs and their primary progenitors
allows us to determine the propagation parameters, such as the 
diffusion coefficient and its energy dependence~[\refcite{Strong2007}].
Among the ratios B/C and Sub-Fe/Fe, it is of great interest to determine the propagation 
history of the lighter nuclei Li, Be, and their isotopes, which are mostly of spallogenic origin.
Their abundances depend not only on interactions of the primary species C, N and O, but 
also on tertiary contributions like Be$\rightarrow$Li or B$\rightarrow$Li. 
Therefore the Li/C and Be/C ratios may provide further restrictions on propagation models~[\refcite{DeNolfo2006}].

AMS-01 observed cosmic rays at $\sim$380 km of altitude during a period 
close to the solar minimum, 
providing data free from atmospheric-induced background
and little influenced by the solar modulation.
Using these data, we present a measurement of the ratio Li/C, Be/C and B/C.

\section{Instrument}    
\label{Sec::Instrument} 
The AMS-01 precursion mission of the Alpha Magnetic Spectrometer (AMS) project~[\refcite{Battiston2010}] 
operated successfully during the 10-day flight STS-91 onboard \textit{Discovery}.

The AMS-01 spectrometer was composed of a cylindrical permanent magnet (analyzing power $BL^2=0.14\,$Tm$^2$), 
a silicon microstrip tracker (six layers of double sided silicon sensors with resolution of 10$\,\mu$m in the bending coordinate), four time of flight (TOF) scintillator planes (time resolution of $\sim$90$\,$psec for $Z>1$ ions)
an aerogel Cerenkov counter (threshold velocity $\beta$=0.985) and anti-coincidence counters~[\refcite{AMS01Report2002}].

The detector response was simulated with \texttt{GEANT3}~[\refcite{GEANT3}]. 
The effects of energy loss, multiple scattering, nuclear interactions and decays
were included in the code, as well as efficiencies, resolutions and reconstruction algorithms. 
Further simulations have been performed 
using \texttt{GEANT4} [\refcite{GEANT4}] and \texttt{FLUKA} [\refcite{FLUKA}] within the framework of 
the \texttt{Virtual MC} tool-kit~[\refcite{VMC}]. 

\section{Data Analysis}   
\label{Sec::DataAnalysis} 
The identification of a cosmic-ray particle with the AMS-01 spectrometer was perfomed
through the combination of independent measurements provided by the various sub-detectors.
The particle rigidity $R$ (momentum per unit charge $p/Z$) was provided by the deflection of the
reconstructed particle trajectory.
The velocity $\beta=v/c$ was measured from the particle transit time between the four TOF
planes along the track length. 
The particle charge magnitude $|Z|$ was obtained by the analysis
of the multiple measurement of energy deposition in the four TOF scintillators (up to $Z=2$~[\refcite{AMS01Report2002}]) 
and the six silicon layers (up to $Z=8$~[\refcite{AMS01Nuclei2010}]). 
Figure~\ref{Fig::ChargeSpectrum} shows the charge histogram of CR particles 
with $Z>2$ obtained with the silicon tracker.
The particle mass was then determined from the resultant charge $Z$, velocity $\beta$ and rigidity $R$:
\begin{figure}
  \centering
  \psfig{file=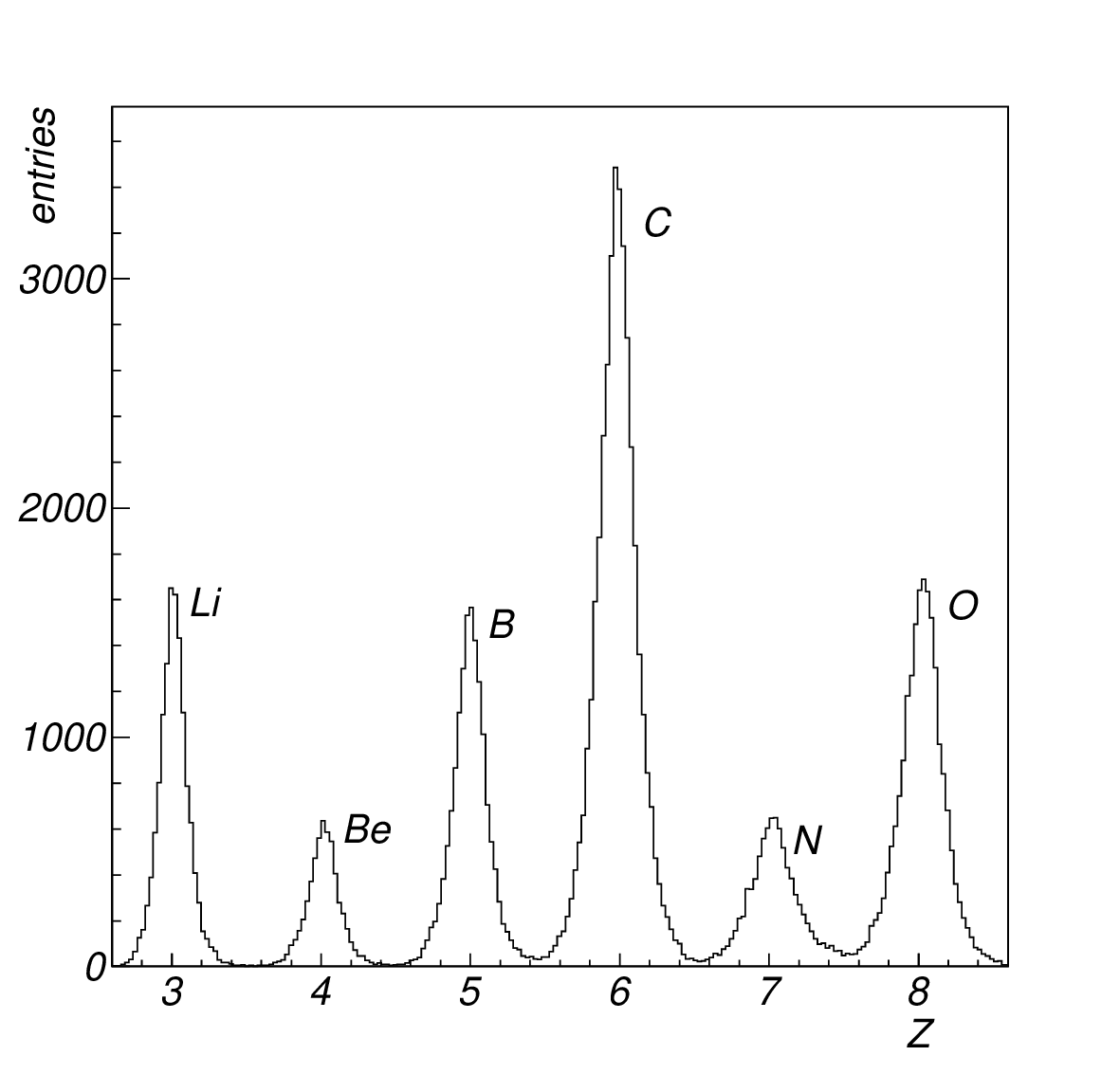,width=3.7in}
  \caption{
    Charge spectrum of the selected $Z>2$ data.  
    The velocity-dependent signal amplitudes of are equalized to $\beta\equiv 1$
    and shown in units of charge. 
    Different nuclear species fall in distinct charge peaks 
    of widths 0.1\--0.16 charge units.
  }
  \label{Fig::ChargeSpectrum}
\end{figure}

The differential energy spectrum of the $Z$-charged particles measured by AMS-01 in
the energy bin $E$ of width $\Delta E$ is related to the measured counts $\Delta N^{Z}$ by:
\begin{equation}\label{EqFluxSimplified}
\Phi^{Z}(E)=\frac{\Delta N^{Z}(E)}{A^{Z}(E)\cdot \Delta T^{Z}(E)\cdot \Delta E}
\end{equation}
where $\Delta T^{Z}(E)$ is the effective exposure time
and $A^{Z}(E)$ is the detector acceptance.
In order to compute bin-to-bin ratios, all the elemental fluxes were determined in a common 
grid of kinetic energy per nucleon $E$, obtained by the rigidity measurements performed by the tracker.
The relation between the measured energy of detected particles and their true energy
were studied using unfolding techniques~[\refcite{DAgostini1995}].
Most of systematic uncertainties arising from many steps of the analysis are suppressed 
in the ratios. Differences in the trigger efficiencies of the two species 
are present, as expected, from the charge dependence of delta ray production and 
fragmentation effects in the detector material. 
This study was performed through extensive simulations employing the 
transport codes \texttt{GEANT3}, \texttt{GEANT4} and \texttt{FLUKA},
Uncertainties of 2\--10~\% were estimated for the trigger efficiency.
The spill over from adjacent charges also produces net effects on the measurements,
leading to errors of up to 5~\%.
The MC determination of the acceptance gave a statistical uncertainty 
of $\sim$1$\--$3$\,$\%, increasing with energy. 

\section{Results and Discussions}  
\label{Sec::ResultsAndDiscussions} 
\begin{figure}
  \centering
  \psfig{file=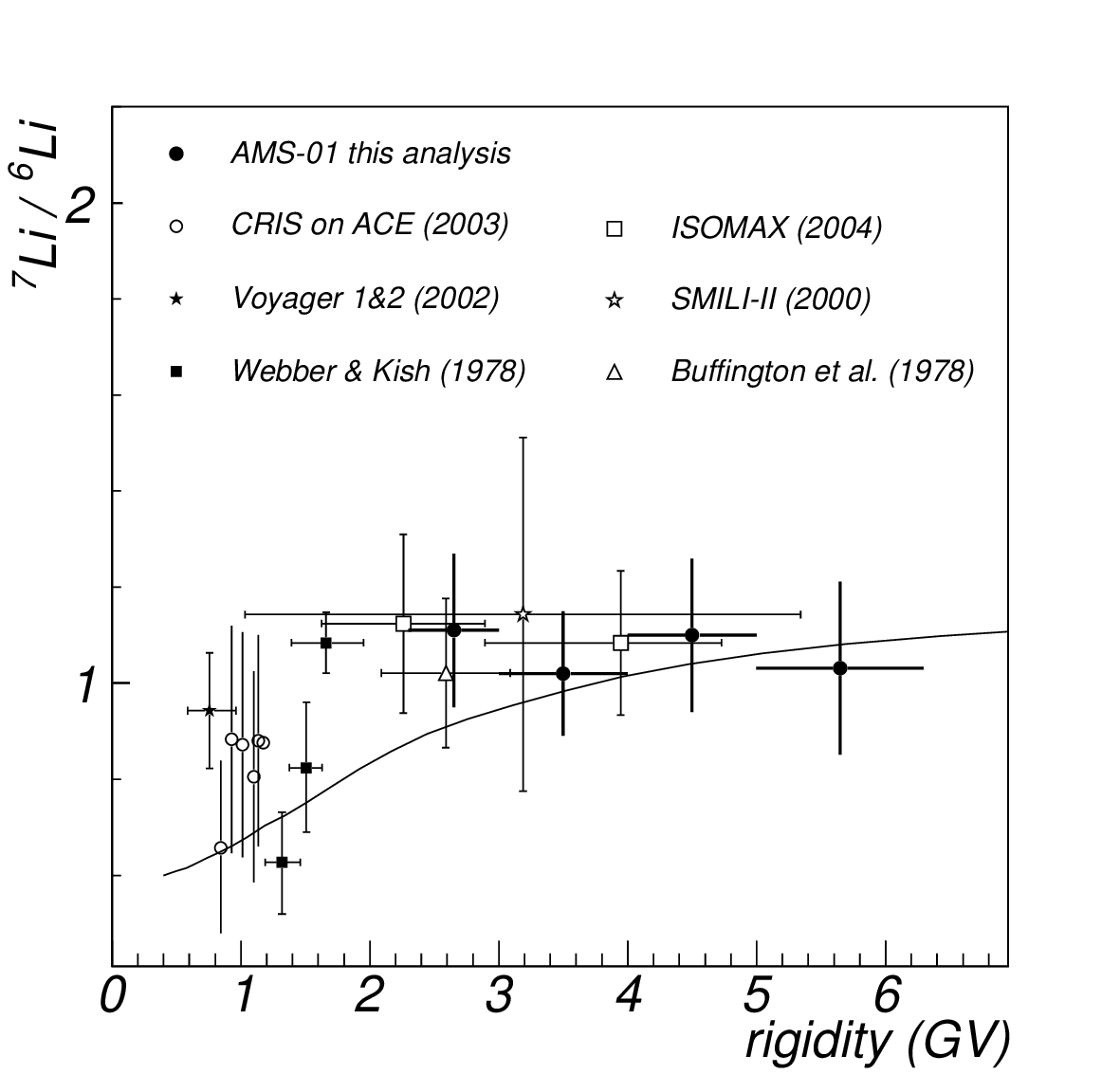,width=3.7in}
  \caption{
    The rigidity dependence of the ratio $^{7}$Li/$^{6}$Li.
    The other experimental values are converted from kinetic energy to rigidity~
    [\refcite{DeNolfo2003,Webber2002,WebberKish1979,Hams2004,Ahlen2000,Buffington1978}].
    The solid line is a diffusion model prediction obtained with \texttt{GALPROP}~(\S\ref{Sec::ResultsAndDiscussions}).  
    \label{Fig::LithiumIsotopicRatio}
  }
  \label{Fig::RatioMassesLiBeB}
\end{figure}

\begin{figure}[!h]
  \centering
  \psfig{file=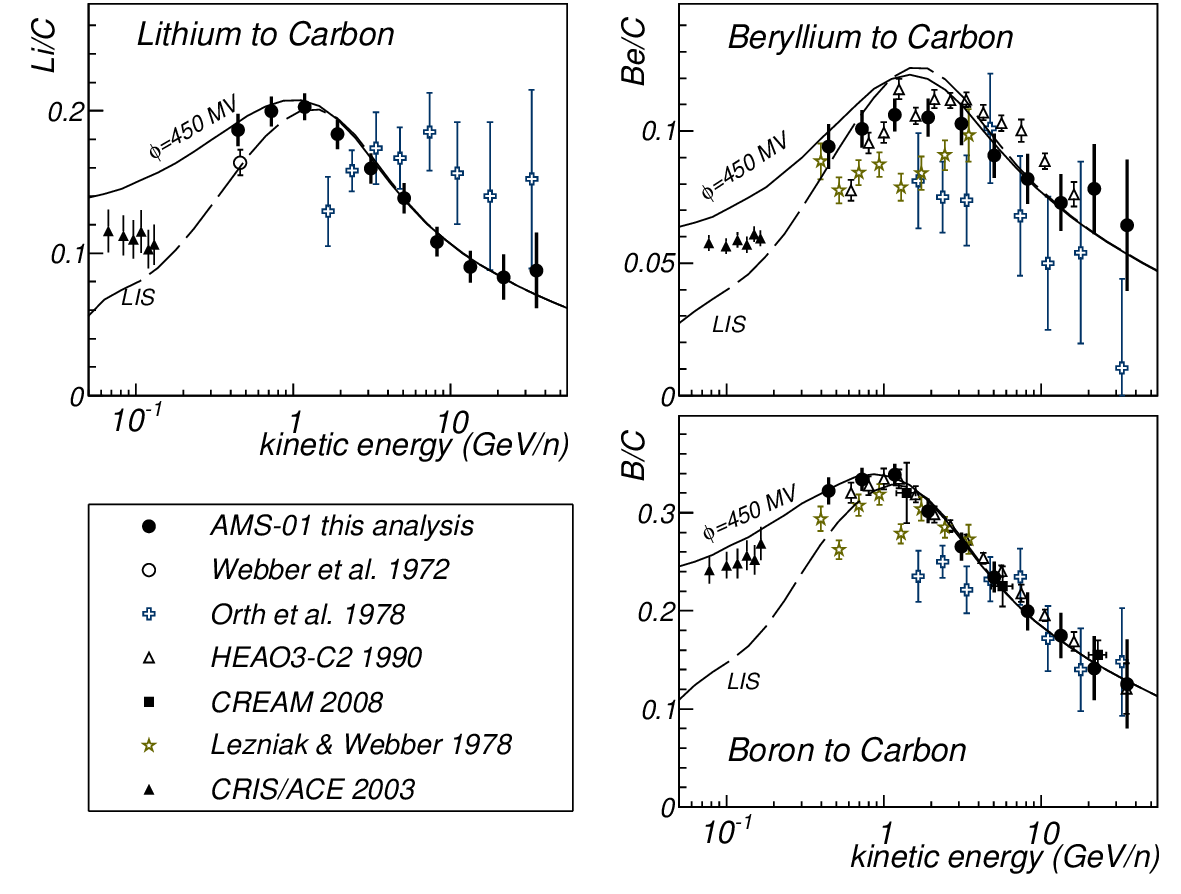,width=4.2in}
  \caption{
    Results for the ratios Li/C, Be/C and B/C from this work (solid circles)
    and previous measurements [\refcite{Webber1972, Orth1978, Lezniak1978, HEAO1990, CREAM2008, DeNolfo2003}]. 
    Model calculations, obtained with \texttt{GALPROP}~\refcite{WebRun2010}, are reported for the interstellar (LIS) 
    and solar modulated ($\phi$=450 MV) cosmic ray fluxes.
  }
  \label{Fig::SecPriRatios}
\end{figure}

Figure~\ref{Fig::LithiumIsotopicRatio} shows the isotopic ratio $^{7}$Li/$^{6}$Li
as a function of the rigidity. Ongoing measurements of other isotopic fractions of H, He, 
Li, Be and B will be discussed during the conference.
Results for the Li/C, Be/C and B/C ratios are presented in Figure~\ref{Fig::SecPriRatios}
with the existing experimental data~[\refcite{Webber1972,Orth1978,Lezniak1978,HEAO1990,CREAM2008,DeNolfo2003}]. 
The error bars in the figure represent the sum in quadrature of 
statistical errors with the systematic uncertainties.

Our B/C ratio measurement agrees well with the results from the first flight of 
CREAM in 2004 [\refcite{CREAM2008}] and with the data collected by HEAO-3-C2~[\refcite{HEAO1990}] 
from October 1979 and June 1980. 
The Be/C ratio is consistent, within errors, with the HEAO data, 
but not with balloon data~[\refcite{Orth1978}].
Our Li/C data have unprecedented accuracy in a poorly explored energy region.
In comparison with balloon data from Ref.~[\refcite{Orth1978}], 
our data indicate a quite different trend in the high energy part of the Li/C ratio. 
In these ratios, the main progenitors of boron nuclei are primary cosmic rays (CNO).
On the contrary, the abundances of Li and Be depend also on secondary progenitors Be and B
through tertiary contributions like B$\rightarrow$Be, Be$\rightarrow$Li and B$\rightarrow$Li.
However, the observed shapes of the measured ratios Li/C and Be/C suggest 
their suitability in constraining the propagation parameters:   
their decreasing behaviour with increasing energy is a direct consequence of the magnetic diffusion
experienced by their progenitors, while the characteristic peak around $\sim$1$\,$GeV/n is 
a strong indicator of low energy phenomena like stochastic reacceleration or 
convective transport with the galactic wind. 

To describe our results, we make use \texttt{GALPROP}\footnote{http://galprop.stanford.edu},  
a numerical code that incorporates all the astrophysical inputs of the CR galactic transport.
In thus proceeding, we use the latext version \texttt{GALPROP-v54} 
through the web-based interface~\texttt{WebRun}~[\refcite{WebRun2010}].
We describe the propagation in the framework of the diffusive-reacceleration model, 
that has been very successful in the description of the cosmic ray nuclei fluxes. 

Our results for the B/C ratio, the Li/C ratio and the $^{7}$Li/$^{6}$Li isotopic ratio
of Figiure~\ref{Fig::LithiumIsotopicRatio} are described quite well within the uncertainties.
It is difficult, however, to accomodate the Li and B description with the Be/C ratio 
by only means of astrophysical parameters.
Beryllium appears to be overproduced in the model by a factor $\sim$10$\--$15$\,$\%. 
This discrepancy is also apparent in the previous measurements.
It should be noted that spallation processes of light nuclei over interstellar 
medium are not well understood quantitatively.
The lack of cross section measurements limits the model predictions 
to uncertainties of $\sim$10--20$\,$\% in the Li/C and Be/C ratios~[\refcite{DeNolfo2006}]. 

Cosmic rays are expected to be measured with high precision in the near future.
Understanding fragmentation is a key factor in establishing
final conclusions concerning cosmic ray propagation.

\bibliographystyle{ws-procs9x6}
\bibliography{ws-pro-sample}

\end{document}